\documentclass[sigplan,10pt]{acmart}
\setcopyright{none}
\renewcommand\footnotetextcopyrightpermission[1]{}
\usepackage{epsfig,endnotes}
\usepackage{lipsum}
\usepackage{url}
\usepackage{xspace}
\usepackage{tabu}
\usepackage{makecell}
\usepackage{xurl}
\usepackage{tabularx}
\usepackage{array,booktabs}
\usepackage[flushleft]{threeparttable}
\usepackage{booktabs}
\usepackage{tikz}
\usepackage{xcolor}
\usepackage{longtable}
\usepackage{enumitem}
\usepackage{amsmath}
\usepackage{delimset}
\usepackage{amsfonts}
\usepackage{color}
\usepackage{listings}
\usepackage{caption}
\usepackage[vlined,ruled,linesnumbered]{algorithm2e}
\usepackage{multirow}
\usepackage{dblfloatfix}
\usepackage[utf8]{inputenc}
\usepackage[T1]{fontenc}
\usepackage[noend]{algpseudocode}
\usepackage[most]{tcolorbox}
\usepackage[normalem]{ulem}
\usepackage{hyperref}
\usepackage{nicefrac}
\usepackage{microtype}
\usepackage{graphicx}
\usepackage{subcaption}
\usepackage{pifont}
\usepackage{float}
\usepackage{textcomp}
\usepackage{cleveref}
\usepackage{arydshln}
\newenvironment{myitemize}%
  {\begin{itemize}
	[leftmargin=0cm,
		itemindent=.3cm,
		labelwidth=\itemindent,
		labelsep=0pt,
		parsep=3pt,
		topsep=2pt,
		itemsep=1pt,
		align=left]
  }%
  {\end{itemize}}

\newenvironment{myenumerate}%
  {\begin{enumerate}
    [leftmargin=*,
        itemindent=0pt,
        labelwidth=.15cm,
        labelsep=.15cm,
        parsep=0pt,
        topsep=2pt,
        itemsep=1pt,
        align=left]
  }%
  {\end{enumerate}}

\usepackage{soul}

\definecolor{dyryu_blue}{rgb}{0.3, 0.7, 1}

\newcommand{\eat}[1]{}

\newcommand{\sys}{\textsf{DIALER}\xspace}
\newcommand{\ours}{\sys}
\newcommand{\oursfull}{\textsf{DIALER} (Directed Inference Adaptation for Local Error Repair)\xspace}
\newcommand{\parlabel}[1]{\vspace{0.2em}\noindent\textbf{#1}}

\renewcommand{\sectionautorefname}{§\kern-2.5pt}
\renewcommand{\subsectionautorefname}{§\kern-2.5pt}
\renewcommand{\subsubsectionautorefname}{§\kern-2.5pt}

\makeatletter
\renewcommand\paragraph{\@startsection{paragraph}{4}{\z@}
  {-.3ex \@plus -.3ex \@minus -.0ex}
  {-1.ex}
  {\normalfont\bfseries}}
\makeatother

\SetAlgorithmName{Algorithm}{Algorithm}

\SetAlFnt{\sffamily}
\SetAlCapFnt{\normalfont\sffamily}

\makeatletter
\renewcommand{\algocf@Vline}[1]{
  \strut\par\nointerlineskip
  \algocf@push{\skiprule}
  \hbox{\bgroup\color{cyan}\vrule\egroup%
    \vtop{\algocf@push{\skiptext}
      \vtop{\algocf@addskiptotal #1}\bgroup\color{cyan}\Hlne\egroup}}\vskip\skiphlne
  \algocf@pop{\skiprule}
  \nointerlineskip}
\renewcommand{\algocf@Vsline}[1]{
  \strut\par\nointerlineskip
  \algocf@bblockcode%
  \algocf@push{\skiprule}
  \hbox{\bgroup\color{cyan}\vrule\egroup
    \vtop{\algocf@push{\skiptext}
      \vtop{\algocf@addskiptotal #1}}}
  \algocf@pop{\skiprule}
  \algocf@eblockcode%
}
\makeatother

\SetKwProg{Fn}{Function}{}{}

\DeclareFixedFont{\ttb}{T1}{txtt}{bx}{n}{8} 
\DeclareFixedFont{\ttm}{T1}{txtt}{m}{n}{8}

\definecolor{codegreen}{rgb}{0,0.6,0}
\definecolor{codegray}{rgb}{0.5,0.5,0.5}
\definecolor{codepurple}{rgb}{0.58,0,0.82}
\definecolor{codecyan}{rgb}{0.0,0.54,0.54}
\definecolor{codeorange}{rgb}{0.9,0.3,0}
\definecolor{backcolour}{rgb}{0.95,0.95,0.92}
\definecolor{deepblue}{rgb}{0,0,0.5}
\definecolor{deepred}{rgb}{0.6,0,0}
\definecolor{deepgreen}{rgb}{0,0.5,0}
\definecolor{gray}{rgb}{0.33,0.33,0.33}

\lstdefinestyle{mystyle}{
    language=Go,
    commentstyle=\color{codegreen},
    keywordstyle=\color{deepblue},
    numberstyle=\tiny,
    basicstyle=\ttfamily\footnotesize,
    breakatwhitespace=false,
    breaklines=true,
    belowskip=0em,
    captionpos=b,
    columns=fullflexible,
    frame=tb,
    keepspaces=true,
    numbers=left,
    numbersep=5pt,
    showspaces=false,
    showstringspaces=false,
    showtabs=false,
    tabsize=2,
    moredelim=**[is][\color{codeorange}]{|}{|},
    escapeinside={<@}{@>},
    moredelim={[is][\color{lightgray}]{@}{@}},
}
\newcommand{\includefigure}[2][]{%
    \IfFileExists{#2}{%
        \includegraphics[#1]{#2}%
    }{%
        \includegraphics[#1]{figures/Dummy Figure.png}%
    }%
}

\begin{document}
\sloppy

\date{}

\title{
DIALER: A Case for Improving Rare-Class Accuracy in Retraining-Free Edge Video Analytics
}

\author{Dongyoon Ryu}
\affiliation{
  \institution{POSTECH}
  \country{}
}
\email{dyryu4w@postech.ac.kr}

\author{Sungho Jeon}
\affiliation{
  \institution{POSTECH}
  \country{}
}
\email{sunghojeon@postech.ac.kr}

\author{Xinyue Ma}
\affiliation{
  \institution{POSTECH}
  \country{}
}
\email{xinyuema@postech.ac.kr}

\author{Di Wang}
\affiliation{
  \institution{}
  \country{}
}
\email{diwangbruce@gmail.com}

\author{Jonghyun Choi}
\affiliation{
  \institution{Seoul National University}
  \country{}
}
\email{jonghyunchoi@snu.ac.kr}

\author{Minjia Zhang}
\affiliation{
  \institution{University of Illinois Urbana-Champaign}
  \country{}
}
\email{minjiaz@illinois.edu}

\author{Myeongjae Jeon}
\affiliation{
  \institution{POSTECH}
  \country{}
}
\email{mj.jeon@postech.ac.kr}

\begin{abstract}

Edge video analytics with lightweight models is prone to accuracy
degradation due to persistent distributional shifts in live video streams.
While continuous learning (CL) addresses such data drift, it heavily
strains the limited compute resources of edge servers originally
provisioned for inference. Our empirical study reveals that emerging vision
foundation models (VFMs) offer a practical, retraining-free alternative
that delivers high average accuracy with remarkable compute savings.
However, VFMs frequently misclassify specific rare classes, which often
represent critical objects, as visually similar common classes. We design
\ours, a system that exploits the spare compute cycles freed by
retraining-free VFM inference to mitigate rare-class misclassifications.
Specifically, \ours pre-builds multi-stage correction pipelines for
dominant rare-to-common confusion pairs offline. At runtime, it routes
correction candidates to the corresponding pipelines and executes as many
stages as idle GPU headroom permits.
Evaluation on four real-world driving datasets shows that \ours improves
rare-class accuracy by up to 14.0\% without interfering with real-time VFM
inference for multi-stream analytics.

\end{abstract}

\maketitle
\pagestyle{plain}
\section{Introduction}

Modern edge video analytics systems typically deploy lightweight deep neural
network (DNN) models (e.g., ResNet18~\cite{resnet}) on GPU servers for tasks
such as object detection and classification~\cite{ekya, recl, ccl}. Located
near the data source (e.g., street cameras), these systems aim to conduct
inference on live video streams in real time~\cite{mark} to extract
actionable insights, such as pedestrian counts or traffic conditions.
However, inference accuracy is frequently compromised by \emph{data drift}, a
persistent distributional shift of live video relative to the training data
(e.g., changes in lighting and weather conditions)~\cite{odin, edgesync,
drift2025coping}. Because lightweight models center on compute efficiency,
they often lack the generalization capability required to handle such
variations, resulting in accuracy drops~\cite{ekya, recl, adainf, dacapo}.

Continuous learning (CL) mitigates the impact of data drift by periodically
retraining models on live video data. While CL allows edge DNNs to maintain
high accuracy, it imposes an insatiable demand on compute resources due to
two preparatory tasks: (1) labeling input samples using heavyweight teacher
models (e.g., WideResNet101~\cite{wideresnet}; 126.9M params) and (2)
searching for retraining hyperparameters, such as batch sizes and training
epochs, through repeated test runs. Together with the retraining process
itself, these tasks consume 3.3$\times$ to 9$\times$ more compute resources
than inference alone~\cite{recl, khani2021real}. This demand strains the
limited compute capacity of edge servers often equipped with only a few
low-power GPUs~\cite{mu, outposts, stackedge}. Existing efforts attempt to
reduce such compute cost by reusing previously trained models for similar
data drifts~\cite{recl} or adopting early-exit model structures to save
inference compute cycles~\cite{adainf}. Unfortunately, these designs fail to
eliminate the dominant cost of retraining and still require complex
co-scheduling of delay-sensitive retraining with live inference.

A model that avoids retraining must satisfy two requirements: it must
generalize well enough to withstand data drift on its own and remain small
enough for real-time edge inference. Emerging vision foundation models (VFMs)
meet both requirements, owing to pretraining techniques
such as distillation that supervises small models (e.g., DINOv3-S; 21M
params) using giant models (e.g., DINOv3-7B; 7B params). Our empirical
evaluation across four real-world driving datasets~\cite{cityscapes, bdd100k,
nuscenes, d2city} reveals that an inference-only workflow with a compact VFM
(DINOv3-S)
outperforms existing retraining-based systems, including Ekya~\cite{ekya},
AdaInf~\cite{adainf}, RECL~\cite{recl}, and MOCHA~\cite{mocha}, by up to 16\%
in accuracy while reducing GPU usage by 47\% to 89\% across 1 to 8 concurrent
video streams.

However, this analysis uncovers a critical blind spot masked by these overall
accuracy gains: \emph{rare classes}, defined by their infrequent occurrence,
are often misclassified as common classes (e.g., 41.0\% of \textsf{Truck} and
23.8\% of \textsf{Bus} samples are misclassified as \textsf{Car}). Although
rare classes contribute little to average accuracy, they frequently
correspond to critical events or objects in real-world
scenarios~\cite{video_anomaly, weapon_detection}. For instance, in driving
video analytics, misclassifying rare large vehicles (e.g., trucks or buses)
as common passenger cars can compromise safety-critical downstream decisions,
such as collision avoidance or lane control.

We present \oursfull{}, a framework that mitigates rare-class misclassifications
during VFM-based inference. \ours{} operates in two phases: \emph{offline
preparation} and \emph{online correction}. During offline preparation,
\ours{} identifies dominant confusion pairs between rare and common classes
through data-driven profiling on curated public datasets~\cite{delta}. It
then clusters each class's samples into fine-grained feature-space groups
using their VFM representations. For each confusion pair, \ours{} builds a
multi-stage correction pipeline per feature-space group. It starts with a
lightweight \emph{adapter} that reuses the representation computed during VFM
inference to fix the easier mispredictions at small compute cost. When the
adapter leaves a sample unfixed, costlier \emph{refinement} stages then
re-examine it. \ours{} therefore spends little compute on the easy correction
cases and reserves the expensive stages for the hard ones. Furthermore, since the rare-class accuracy
gains accumulate, the pipeline naturally exposes a sequence of operating
points at which correction can stop.

At runtime, \ours{} uses these pipelines to verify whether a low-confidence
common-class prediction by the VFM is valid or a false positive that requires
correction. To avoid delaying real-time inference, the correction phase
harnesses spare GPU time and memory left available after VFM inference.
To operate within this budget, \ours{} incorporates a \emph{budget-aware
scheduler} that tracks the remaining budget in each inference window and
prioritizes correction work while adapting the amount of additional
processing accordingly. When resources become constrained, \ours{} preserves
the adapter and progressively drops refinement stages that can no longer fit
within the serving deadline. Importantly, \ours{} also bounds GPU memory
overhead by limiting the correction models resident on the GPU and
overlapping model loading with ongoing correction work.

Compared to a baseline training-free VFM setup, our case study demonstrates
that \ours improves rare-class accuracy by up to 14.0\% across four
real-world driving datasets, including Cityscapes~\cite{cityscapes},
BDD100K~\cite{bdd100k}, NuImages~\cite{nuscenes}, and D2City~\cite{d2city},
while degrading common-class accuracy by less than 1\%. This efficacy is
validated on an NVIDIA RTX 3090 GPU and a Jetson Orin
NX using three compact VFM backbones: DINOv3-S (21M
params)~\cite{dino_v3}, SigLIP2-B (86M params)~\cite{siglip2}, and MAWS-B
(86M params)~\cite{MAWS}. 
Furthermore, \ours significantly outperforms
alternative retraining-free designs, such as feature-space methods (prototype
classifiers~\cite{prototype_classifier} and kNN~\cite{feature_space_knn}) and a
shared MLP~\cite{shared_mlp},
which yield modest rare-class gains of only 3.4\%, 4.5\%, and 6.0\%, respectively,
compared to 14.0\% in \ours.

In summary, we make the following key contributions:

\begin{myitemize}
\item We provide the first systematic study revisiting retraining paradigms
for edge VFMs.
Our key findings reveal that while full model fine-tuning is no longer necessary
to achieve high aggregate accuracy, severe rare-class accuracy degradation persists
behind these statistical averages.

\item We present \ours, a retraining-free framework that pairs offline
feature profiling with online selective correction for rare classes.
\ours operates in an idle-resource-aware manner, exploiting
opportunistic GPU headroom without affecting primary VFM inference latency.

\item Through extensive evaluations on edge platforms, we show that \ours
improves rare-class accuracy by up to 14.0\% with negligible
common-class degradation, significantly outperforming existing
feature-space and adaptation baselines.
\end{myitemize}

\section{Background and Motivation}
\label{sec:background}

This section presents the background on live edge video analytics and discusses 
the key motivations behind our study.

\subsection{Video Analytics at the Edge}
\label{sec:va_edge}

\parlabel{Common Practices.} While public clouds offer vast compute
capacity, edge servers are the primary deployment target for live video
analytics due to two critical factors: privacy regulations and network
overheads. First, live video streams captured at the edge (e.g., surveillance
and in-vehicle cameras) frequently record user-sensitive or personal data.
Strict data privacy regulations, such as GDPR~\cite{gdpr} and CCPA~\cite{ccpa},
strongly favor processing and storing such content locally. Second, video
streaming is inherently bandwidth-intensive~\cite{videostorm, scaling}; for
instance, a 4K security camera capturing compressed video at 30 frames per
second (FPS) can generate data rates of up to 56~Mbps~\cite{bitrate}. Given
that edge locations frequently face constrained or volatile uplink
bandwidth, continuous cloud offloading is both cost-prohibitive and unreliable. 
Consequently, modern video analytics systems execute
end-to-end inference pipelines directly on edge
nodes~\cite{axiom_vision, delta}.

At the core of video stream analysis is model inference, where DNNs process
incoming frames for visual tasks such as object
detection, classification, and tracking. To match high frame rates, these
inference tasks must satisfy strict, millisecond-level service level
objectives (SLOs)~\cite{adainf}. To meet these tight latency bounds on edge
servers that usually host only a few energy-efficient GPUs (e.g., NVIDIA
A2~\cite{nvidia_a2})~\cite{mu, outposts, stackedge}, prior systems have
commonly relied on lightweight convolutional neural networks (CNNs), such as
ResNet~\cite{resnet}, YOLO~\cite{yolo}, and ENet~\cite{enet}, which feature
compact architectures with fewer parameters.

\parlabel{Data Drift.} Live video streams at inference time may diverge from
the original training data as objects and scenes inevitably evolve over
time~\cite{odin, edgesync, drift2025coping}. This phenomenon, termed
\emph{data drift}, poses a key challenge to edge video analytics. Data drift
can arise from various sources, including environmental conditions (e.g.,
lighting and weather), camera setups (e.g., angle and resolution), and object
appearances (e.g., vehicle designs and road hazards). For example, driving
video analytics systems can face drastic illumination shifts across weather
and time-of-day variations, as well as dynamic scene changes in urban traffic
environments~\cite{ekya}. Although edge models satisfy latency SLOs, their
limited model capacity renders them vulnerable to data drift, causing severe
performance degradation.

\parlabel{Continuous Learning Pipelines.} CL has
emerged as a key paradigm for mitigating data drift in edge video
analytics~\cite{ekya, dacapo, adainf, eomu, axiom_vision, ccl}. By
periodically retraining on a subset of streaming data, CL adapts the deployed
model to evolving data distributions. A CL pipeline typically comprises three
core stages:

\begin{myitemize}

    \item \emph{Labeling} employs a large and compute-expensive
    teacher model (e.g., ResNeXt101~\cite{resnext} or WideResNet101~\cite{wideresnet})
    to annotate a small fraction of incoming frames.
    This automated pseudo-labeling eliminates manual annotation effort.

    \item \emph{Profiling} balances the expected accuracy gain of retraining
    against its compute cost.
    It searches retraining hyperparameters (e.g., training epochs,
    batch size, and frozen layers) through fast trial runs on the labeled
    samples and estimates the impact of drift on each
    stream to decide which stream to retrain first.

    \item \emph{Retraining} fine-tunes the deployed edge model using the newly labeled samples and the selected hyperparameter configuration to restore inference accuracy amid data drift.
\end{myitemize}

Two primary challenges impede the efficient execution of 
CL tasks alongside live inference on a shared edge server.  First,
retraining is triggered either at fixed intervals~\cite{ekya} or reactively
upon drift detection~\cite{dacapo, eomu}, each incurring distinct drawbacks:
periodic retraining wastes scarce compute when drift is minimal, whereas
reactive drift detection introduces non-trivial monitoring complexity and
overhead. Second, existing CL tasks collectively dominate the temporal and
spatial usage of GPU compute resources
(\autoref{fig:resource_usage})~\cite{recl, khani2021real}. When co-located
with live inference, this heavy compute footprint incurs severe resource
contention, which not only creates huge lags in model adaptation but also
risks violating inference SLOs.

\subsection{Rethinking Accuracy}
\label{sec:rethinking_acc}

We now highlight two under-explored dimensions in edge video analytics system
design that motivate our empirical study in~\autoref{sec:empirical-studies}.

\begin{figure}[t]
  \centering
  \includefigure[width=\columnwidth]
  {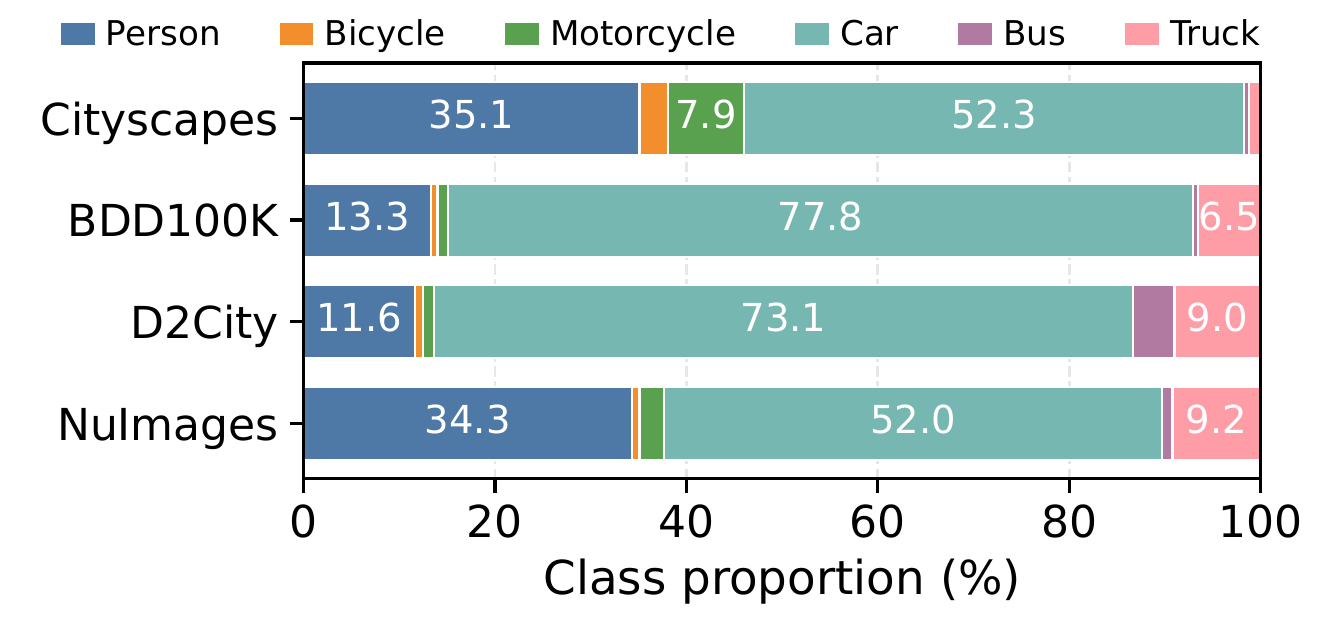}
  \caption{Class composition of annotated objects in the four deployment streams.
  \textsf{Car} and \textsf{Person} dominate every stream, while each of
  the four rare classes accounts for under 10\%.}
  \label{fig:class_composition}
\end{figure}

\begin{myitemize}

\item \textbf{Model Dimension: VFMs vs. CNNs.} Unlike CNNs, which
    primarily capture local features through hierarchical convolutional
    layers~\cite{ma2015hierarchical}, VFMs
    typically leverage vision transformers as backbones. By utilizing
    self-attention mechanisms, such architectures model long-range
    dependencies across image patches and enable a holistic understanding of
    global context in contrast to the restricted local receptive fields of CNNs.
    Furthermore, recent VFM variants include compact versions with
    5--25M parameters 
    (e.g., EVA-02~\cite{eva02}, DINOv3~\cite{dino_v3}, TIPS~\cite{tips}),
    making real-time VFM inference increasingly viable on edge
    hardware. Among the candidate models at
    hand, DINOv3~\cite{dino_v3} stands out for its superior frozen-feature
    generalization, achieved through pre-training on LVD-1689M that comprises
    ${\sim}$1.7B curated images---far exceeding the scale of standard
    datasets like ImageNet-21k ($\sim$14M images). Crucially, such robust representations
    allow frozen VFM backbones to yield high-quality features for downstream visual
    tasks without task-specific retraining.

\item \textbf{Metric Dimension: Per-Class Accuracy Gap.} Class imbalance
    naturally arises in video streams when certain \emph{common classes}
    significantly outnumber \emph{rare classes}. Such class imbalance is
    prevalent in real-world driving streams, as shown in~\autoref{fig:class_composition}.
    In video analytics, rare
    classes frequently correspond to infrequent yet safety-critical objects or
    events that demand highly accurate detection~\cite{video_anomaly}.
    For instance, failing to detect wild animals entering a
    highway poses severe accident risks~\cite{wild_animal,
    saxena2021animal}. Similarly, in driving streams, misclassifying
    rare large vehicles (e.g., trucks) as common passenger cars can
    compromise downstream functions such as trajectory planning and collision avoidance.

    Most retraining-based systems optimize for average accuracy
    across all classes~\cite{ekya, dacapo, adainf}. However, models
    trained on class-imbalanced datasets tend to bias their predictions toward common
    classes~\cite{he2009learning}. This bias leaves rare classes highly susceptible
    to misclassification, particularly when they share visual features with
    common classes.

\end{myitemize}

The above discussion raises two research questions:

\begin{myenumerate}

    \item Can the robust generalization of VFMs obviate the need for
    resource-intensive CL in edge video analytics? If so, how does
    retraining-free VFM inference compare with contemporary
    retraining-based CL systems in terms of accuracy and resource
    efficiency?

    \item If retraining-free VFM inference proves competitive, does it
    achieve balanced per-class accuracy across both common and rare
    classes?
    
\end{myenumerate}

\section{Empirical Studies}
\label{sec:empirical-studies}

In this section, we investigate the questions raised
in~\autoref{sec:rethinking_acc} through a series of experiments across four
datasets and two model architectures.

\subsection{Experimental Setup}
\label{sec:empirical-setup}

\parlabel{Datasets and Streams.} Although our evaluation later extends to
object detection, we focus here on classification tasks across four
real-world driving datasets: 
Cityscapes~\cite{cityscapes} ($\sim$25K European urban scene frames),
BDD100K~\cite{bdd100k} ($\sim$100K 40-second driving videos),
NuImages~\cite{nuscenes} ($\sim$93K driving frames), and
D2City~\cite{d2city} ($\sim$10K dashcam video clips of diverse traffic
scenarios). 
Collectively, these datasets differ in scale and class composition while
covering a wide range of geographies, times of day, weather conditions, and traffic environments, capturing realistic spatio-temporal data drift.

Following prior work~\cite{ekya, dacapo}, we crop individual objects from
video frames and map their annotations to six classes: \textsf{Person},
\textsf{Bicycle}, \textsf{Motorcycle}, \textsf{Car}, \textsf{Bus}, and
\textsf{Truck}. We preserve the temporal order of these cropped objects
to construct a sequential classification stream for each dataset. Similar to
RECL, which uses an initial temporal portion of a video stream for
pre-deployment training~\cite{recl}, we divide each stream into disjoint
offline and online partitions: the first 50\% is used for pre-deployment
model preparation, while the remaining 50\% is reserved for online stream
replay and evaluation. To emulate a multi-camera edge setup, we treat each
dataset stream as an independent workload and co-locate multiple streams on a
shared GPU~\cite{ekya, videostorm, gemel}.

As summarized in~\autoref{fig:class_composition}, \textsf{Car} dominates the
deployment streams (52.0--77.8\%), followed by \textsf{Person}
(11.6--35.1\%), while the remaining four classes each account for under 10\%
(and typically below 5\%). In the remainder of this paper, we define
\textsf{Car} and \textsf{Person} as the common classes and \textsf{Bicycle},
\textsf{Motorcycle}, \textsf{Bus}, and \textsf{Truck} as the rare classes.

\parlabel{Models.} For the retraining baselines, we use
ResNet18~\cite{resnet} as the student and ResNeXt101~\cite{resnext} as the
teacher, following prior work~\cite{ekya, eomu}. For the retraining-free VFM
approach (referred to as \textsf{Frozen}), we primarily use
DINOv3-S~\cite{dino_v3}, one of the three VFMs evaluated in our study.
DINOv3-S is pretrained on $\sim$1.7B images using DINOv3's self-supervised
learning and distillation pipeline, and its weights remain fixed throughout
online serving.

\parlabel{CL Baselines.} We compare \textsf{Frozen} against four
state-of-the-art retraining-based CL systems: Ekya~\cite{ekya},
AdaInf~\cite{adainf}, RECL~\cite{recl}, and MOCHA~\cite{mocha}. Ekya triggers
retraining at fixed intervals and employs a priority-based scheduler to
allocate GPU resources across streams based on profiled accuracy gains.
AdaInf reactively detects data drift via runtime accuracy drops and applies
incremental retraining together with early-exit inference to reduce
retraining interference while satisfying inference SLOs. Meanwhile, RECL
mitigates retraining costs by maintaining a repository of historical expert
models and reusing the one that best matches the current data distribution.
MOCHA extends model reuse by organizing historical experts into a
hierarchical taxonomy of scene semantics and maintaining an on-device cache
to minimize retrieval and switching overheads. For a fair comparison, we
evaluate all baselines under the same GPU compute budget: CL
tasks that run remotely or on shared resources in the original systems
(e.g., MOCHA's cloud tier) are executed on the same local server, while their
model-selection and retraining policies are preserved.

\parlabel{Evaluation Protocol and Implementation.} All experiments in this
section run on an NVIDIA RTX 3090 GPU (24\,GB VRAM) with PyTorch 2.11~\cite{pytorch}. We
leverage CUDA Multi-Process Service (MPS)~\cite{nvidia_mps} for GPU partitioning across
concurrent streams and orchestrate workloads using Ray 2.56~\cite{ray}. Each stream is
replayed at 60 samples/s.
For live inference, both VFM and retraining baselines use a batch size of 60, whereas
retraining batch sizes are dynamically tuned by the CL profiling
task~\cite{ekya}.

\parlabel{Metrics.} To evaluate both model performance and system-level
efficiency, we measure (1) average accuracy across all samples, (2) per-class
accuracy,
(3) GPU-time share (\%) of each task (i.e., inference, labeling, profiling,
model selection, and training), computed by integrating GPU utilization per
task over time, and (4) GPU memory footprint (GB) allocated across active
processes.

\subsection{Accuracy and GPU Resource Usage}
\label{sec:accuracy_and_gpu_resource_usage}

We examine how the evaluated systems trade online GPU resources for accuracy under increasing stream concurrency. We co-locate one, two, four, and eight streams on the same fixed GPU capacity, forcing inference and CL tasks to compete for progressively tighter resources.

\begin{figure}[t]
  \centering
  \includefigure[width=\columnwidth] {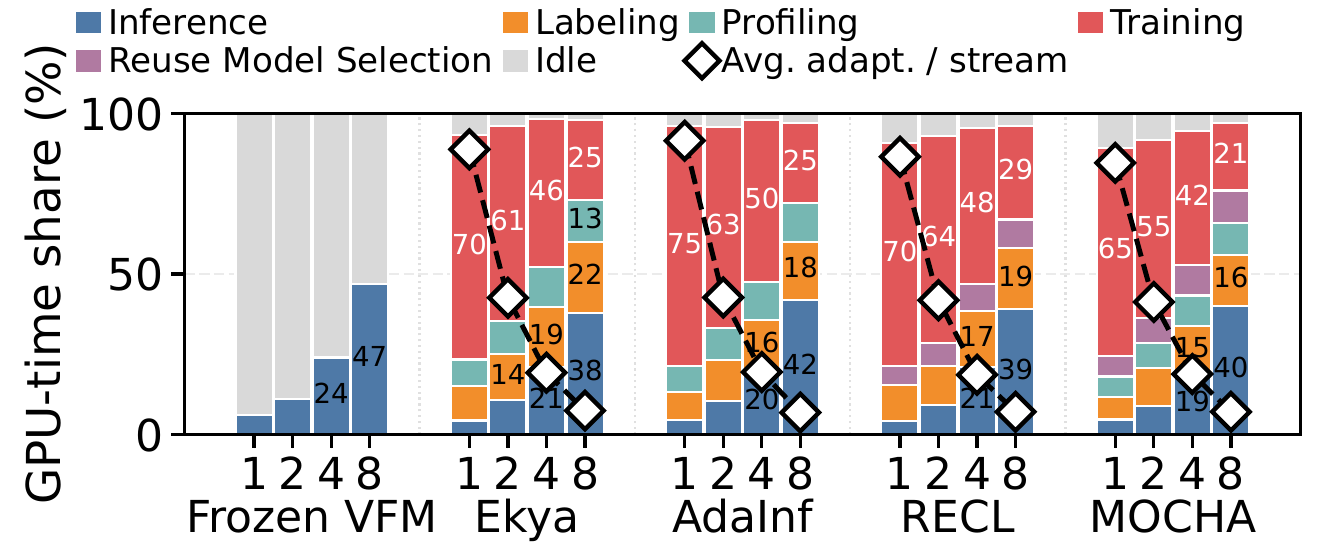}
  \caption{
    Breakdown of GPU-time share by task with single- to multi-stream serving.
    Diamonds mark the average share per stream for CL tasks.
    As inference demand grows with concurrency, CL systems retain
    progressively less GPU time for CL tasks. \textsf{Frozen} spends GPU time
    on inference alone.
  }
  \label{fig:resource_usage}
\end{figure}

\parlabel{Resource Usage.} We first examine how the evaluated systems consume
the fixed GPU capacity as stream concurrency increases.
\autoref{fig:resource_usage} breaks down GPU usage across inference and
CL tasks, including labeling, profiling, model selection, and retraining.
Each additional stream introduces another inference workload, which leaves
less capacity for adaptation to data drift.
CL systems distribute this remaining capacity across streams according to
their own scheduling policies. Consequently,
the GPU capacity available for CL tasks per stream decreases on average as
concurrency increases (diamonds in \autoref{fig:resource_usage}). In
contrast, \textsf{Frozen} executes only the fixed inference path
and uses only 6\% to 47\% of the GPU time, whereas the
CL systems consume nearly the entire GPU capacity.
This overhead extends to GPU memory. Under single-stream serving, \textsf{Frozen} keeps only the inference stack, while CL systems maintain additional GPU state for adaptation tasks. \textsf{Frozen} uses 1.5\,GB, compared with 3.7\,GB for Ekya, 4.2\,GB for AdaInf, 4.9\,GB for RECL, and 4.8\,GB for MOCHA.

\begin{figure}[t]
  \centering
  \includefigure[width=\columnwidth]{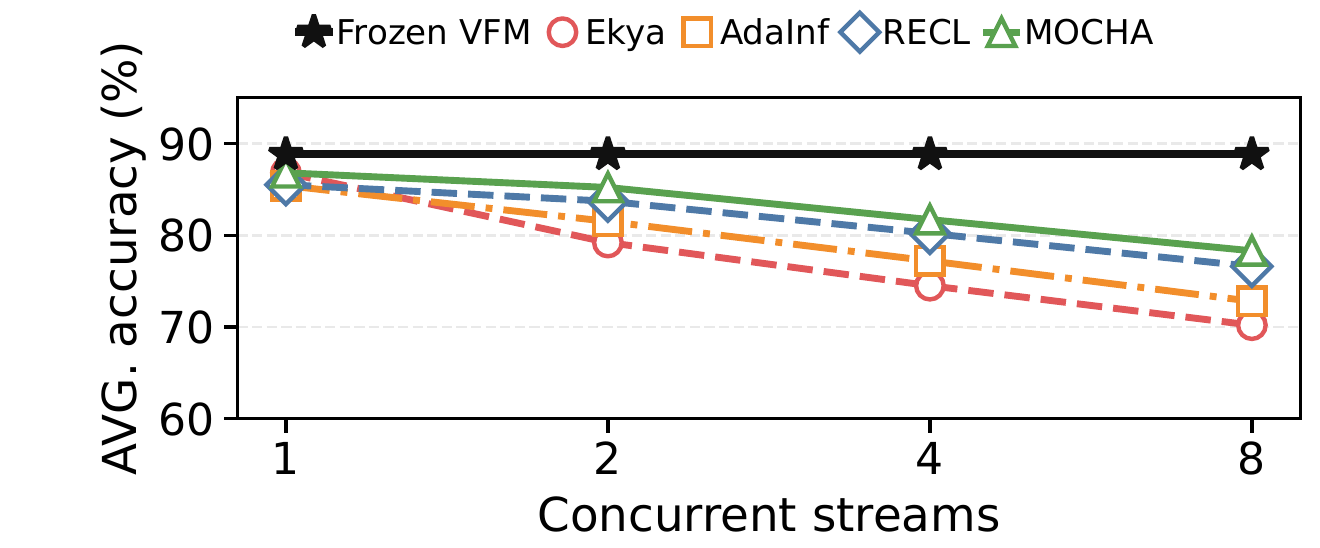}
  \caption{
    Average accuracy for each system while scaling the number of streams. As the number of streams increases, the retraining baselines allocate fewer resources for retraining, resulting in lower accuracy, whereas \textsf{Frozen} achieves favorable accuracy.
  }
  \label{fig:resource_tradeoff}
\end{figure}

\parlabel{Accuracy.} We next examine how these resource behaviors translate
into prediction accuracy. \autoref{fig:resource_tradeoff} reports the average
accuracy under the same four deployment configurations. The CL baselines
lose accuracy as inference consumes a larger fraction of the GPU and less
capacity remains for retraining. Ekya and AdaInf degrade more sharply with
increasing concurrency, primarily because they lack the expert model reuse
that RECL and MOCHA rely on. Even RECL and MOCHA lose accuracy as serving
load grows, since expert selection and retraining both compete for the scarce
GPU headroom left after inference.

In contrast, \textsf{Frozen} achieves 88.8\% average accuracy
at all concurrency levels without any online labeling, profiling, or
retraining. This result shows that a compact VFM alone can provide a strong
deployment point for edge video analytics. However, given the highly skewed
class distribution of our deployment streams, this aggregate accuracy may not
reflect prediction quality for rare classes. We therefore next examine the
per-class accuracy of \textsf{Frozen}.

\begin{figure}[t]
\centering
\includefigure[width=\linewidth]{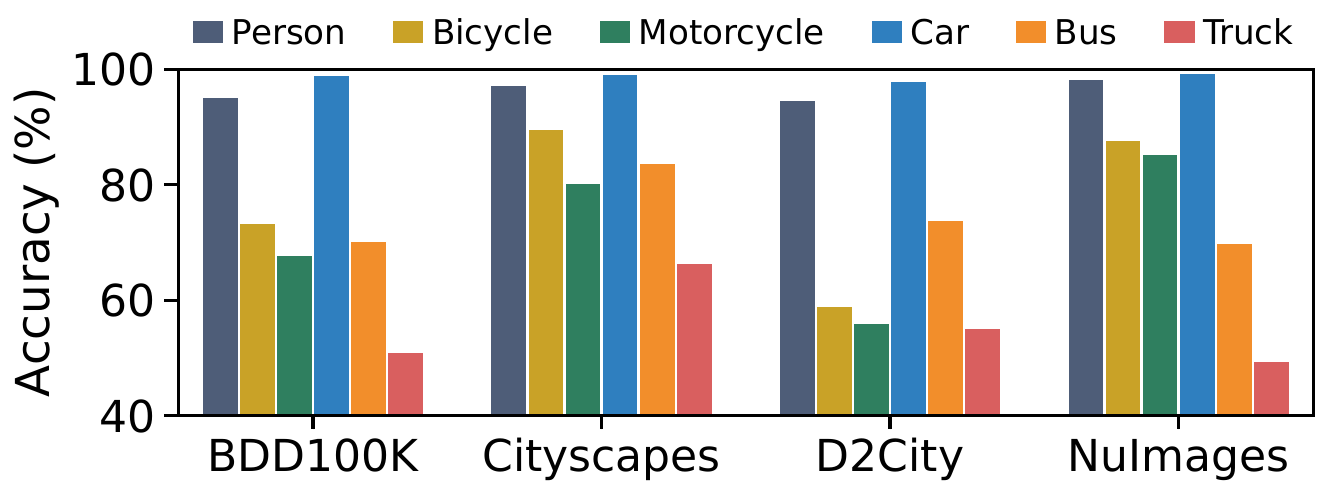}
\caption{
  Per-class accuracy of \textsf{Frozen} on the four deployment streams.
  The common classes (\textsf{Person} and \textsf{Car}) achieve higher
  accuracy than the four rare classes on every stream.
}
\label{fig:per-class_accuracy}
\end{figure}

\begin{figure}[t]
\centering
\includefigure[width=\linewidth]{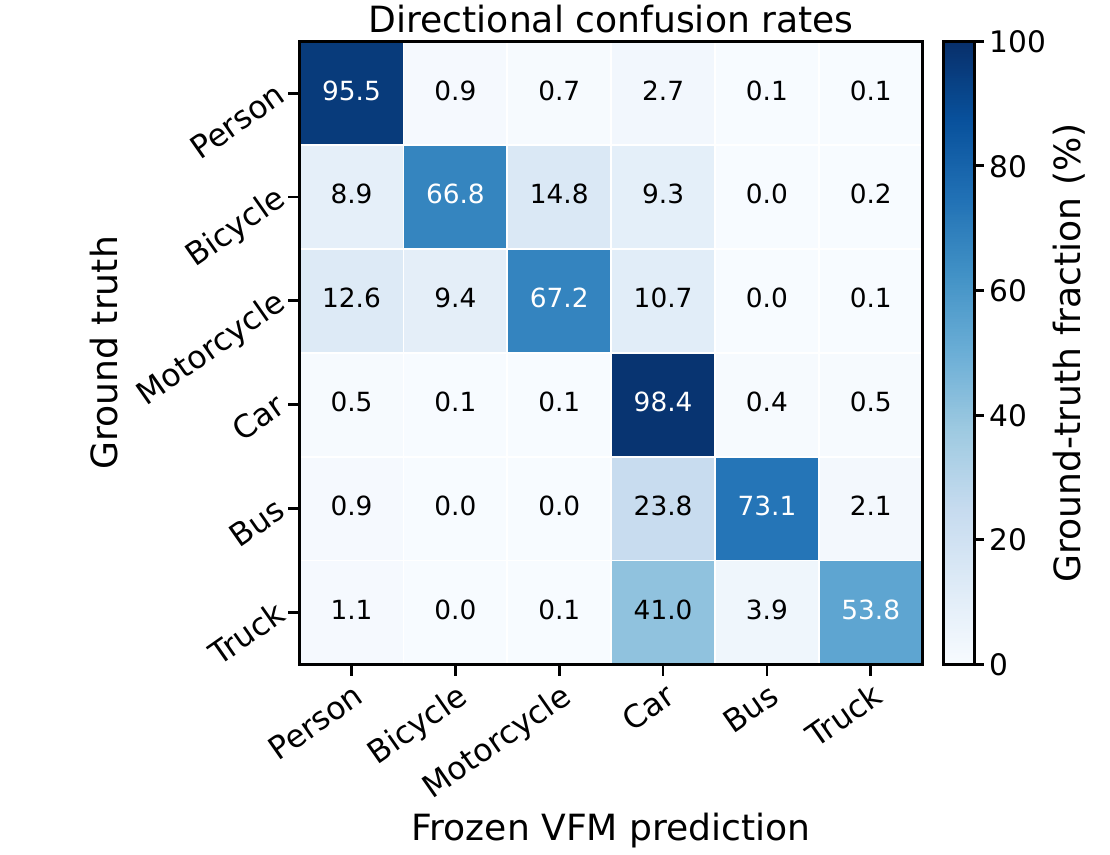}
\caption{
Row-normalized confusion matrix of the frozen VFM. Each row shows the
prediction distribution for a ground-truth class. Rare classes exhibit
substantial confusion toward the two common classes, reaching 18--42\% per
class.
}
\label{fig:confusion_flow}
\end{figure}

\subsection{Structured Rare-Class Confusions}
\label{sec:empirical-confusion}

Because \textsf{Frozen}'s accuracy barely changes with concurrency
(\autoref{fig:resource_tradeoff}), we analyze its per-class behavior in the
single-stream setting. Although \textsf{Frozen} reaches 88.8\% overall accuracy,
~\autoref{fig:per-class_accuracy} shows that this aggregate masks a consistent
gap between common and rare classes across all four deployment streams. To
characterize the structure of these rare-class errors, we next examine the
row-normalized confusion matrix (\autoref{fig:confusion_flow}). Accuracy
remains high for the two common classes, reaching 95.5\% for \textsf{Person}
and 98.4\% for \textsf{Car}, while all four rare classes are considerably
less reliable: \textsf{Bicycle}, \textsf{Motorcycle}, \textsf{Bus}, and
\textsf{Truck} achieve only 66.8\%, 67.2\%, 73.1\%, and 53.8\%, respectively.

More importantly, misclassifications are not distributed uniformly across
classes but concentrate in a small number of dominant confusion pairs, as
shown in the row-normalized confusion matrix (\autoref{fig:confusion_flow}).
The most pronounced examples are \textsf{Truck}$\rightarrow$\textsf{Car} and
\textsf{Bus}$\rightarrow$\textsf{Car}, which account for 41.0\% and 23.8\% of
their respective ground-truth classes. The reverse confusions are much
smaller: only 0.5\% of \textsf{Car} samples are predicted as \textsf{Truck},
and 0.4\% as \textsf{Bus}. Other rare classes exhibit different pairwise
confusion structures. For example, 14.8\% of \textsf{Bicycle} samples are
predicted as \textsf{Motorcycle}, while \textsf{Motorcycle} is most
frequently confused with \textsf{Person} (12.6\%) and \textsf{Car}
(10.7\%). These results show that the misclassifications of \textsf{Frozen}
are structured around a few confusion pairs rather than being diffuse across
all labels.

\parlabel{Observations.} Taken together, our empirical analysis shows that
\textsf{Frozen} achieves strong aggregate accuracy at low GPU cost without
online retraining, but its rare-class accuracy remains low and its errors
concentrate in a few dominant confusion pairs. Because most of its
predictions are already correct, improving rare-class accuracy need not
involve replacing or continuously retraining the entire inference model.
Instead, a finer-grained approach that preserves the VFM inference path and
spends additional computation only on predictions likely to fall into these
confusion pairs can recover much of the lost rare-class accuracy at a
fraction of the retraining cost. The next section examines the opportunities
and system challenges in realizing such selective correction.

\section{Opportunities and Challenges}
\label{sec:opportunities}

In this section, we discuss key opportunities and challenges in enhancing
rare-class accuracy.

\parlabel{Opportunities.} Deploying a frozen VFM removes the GPU overhead of
CL tasks, which creates two opportunities:

\begin{myitemize}

    \item \textbf{Reclaimed Serving Resources:}
    CL shares the edge GPU between inference and CL tasks, whereas frozen VFM inference leaves that capacity available for additional processing during serving (\autoref{fig:resource_usage}).
    Because VFM inference runs within a predictable latency envelope, the slack in each inference window is known at runtime and can be spent on auxiliary work without violating the serving deadline.
    
    \item \textbf{Decoupling Rare-Class Correction from Live Streams:}
    CL depends on incoming frames for retraining, where rare objects appear
    too infrequently to provide effective supervision.
    Curated public repositories instead offer rich samples of rare
    classes~\cite{delta}, and the frozen VFM backbone maps both these
    samples and live frames into the same embedding space.
    Correction models for the dominant confusion pairs can therefore be
    prepared offline.

\end{myitemize}

\parlabel{Challenges.} Exploiting these opportunities on resource-constrained
edge servers introduces three challenges:

\begin{myitemize}
    \item \textbf{Identifying Correction Candidates:}
    The confusion pairs profiled offline do not reveal whether an
    individual online prediction is wrong.
    Applying correction indiscriminately would waste resources on
    already-correct predictions and may flip correct common-class
    predictions.
    The system must therefore decide, using only information available at
    inference time, which predictions warrant correction.

    \item \textbf{Ordering Correction by Expected Gain:}
    Correction can proceed at different depths, and the available budget
    determines where it must stop.
    Correction is therefore useful only if it is ordered so that the
    earliest and cheapest work yields the largest accuracy gain and later
    work adds progressively less.
    Otherwise, a sample whose budget runs out may gain little accuracy from
    the GPU time already spent on it.

    \item \textbf{Satisfying Latency SLO and Memory Constraints:}
    Although the slack in each inference window is known at runtime, the
    cost of a correction is not: loading a non-resident model consumes both time and memory. The two budgets are therefore coupled, and how much of each remains changes as correction proceeds and as more streams share the server. The system must track both budgets and predict the cost of each
    correction against them before dispatching it.    
\end{myitemize}

These opportunities and challenges motivate \ours{}, which uses resources
left by frozen VFM inference to improve rare-class accuracy.

\begin{figure*}[t]
\centering
\includefigure[width=\linewidth]{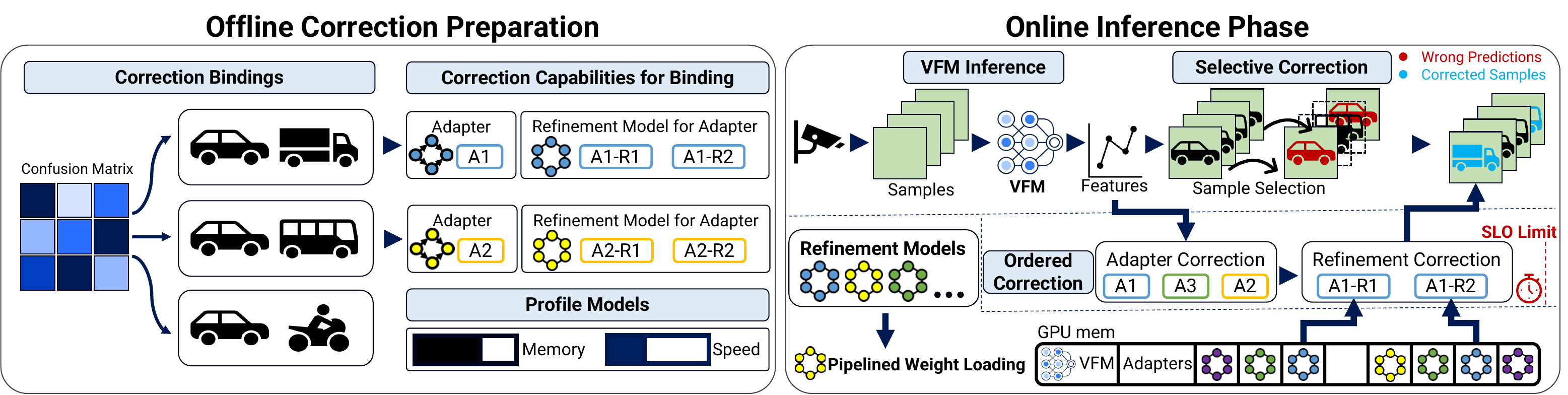}
\caption{
Overview of \ours{}. \emph{Offline:} \ours{} constructs correction \emph{Bindings} from profiled confusion patterns, trains \emph{Binding}-specific adapters and refinement models, and profiles their execution time and memory cost. \emph{Online:} \ours{} reuses cached VFM features to apply ordered correction to selected samples, overlaps refinement-model loading with correction execution, and stops optional work before the serving deadline.
}
\label{fig:system-overview}
\end{figure*}

\section{\ours{}: System Design}
\label{sec:system-design}

\ours{} uses the resource slack left by VFM inference to correct recurring rare-class errors. To support these runtime decisions efficiently, \ours{} prepares reusable correction capabilities and profiles their decision behavior and execution costs to guide runtime selection and execution. At runtime, \ours{} starts from the original VFM predictions as \emph{Base} predictions and selectively applies additional correction where further processing is admitted. The amount of correction is adapted to the resource budget of each inference window, allowing different samples to receive different levels of correction while preserving the latency and throughput objectives of the serving pipeline.

\subsection{Design Goals and Overview}
\label{sec:system-overview}

\parlabel{Design Goals.}
\ours{} is designed to improve rare-class accuracy using resources that remain available during serving. This objective creates three requirements. 
First, correction must be selective to avoid unnecessary changes to high-confidence \emph{Base} predictions.
Second, correction should support graduated levels of effort, beginning with lightweight correction and allowing selected decisions to receive additional refinement. 
Third, correction must preserve the end-to-end serving SLO~\cite{mark}. Correction must therefore account for the time remaining before the inference-window deadline and the GPU memory available for correction.

\parlabel{Two-Phase Design.}
Preparing specialized correction capabilities requires labeled data, training, and profiling, whereas serving must select among them within the current resource budget. \ours{} therefore separates these responsibilities into an offline preparation phase and an online correction phase. 
\autoref{fig:system-overview} shows the resulting design. 
In the offline phase, \ours{} uses labeled data to identify recurring
correction targets and their feature-space contexts, prepare the corresponding correction capabilities, and derive runtime decision criteria and execution profiles.
In the online phase, \ours{} starts from the \emph{Base} predictions,
identifies relevant correction candidates, and selectively executes the
prepared capabilities within the remaining time and GPU memory.

\subsection{Offline Correction Preparation}
\label{sec:offline-correction}
The offline phase prepares reusable correction capabilities from labeled
data~\cite{delta} before deployment. \ours{} first runs the deployed VFM and shared
classifier on these samples and clusters their VFM representations to define
feature-space regions for specialization. Confusion patterns observed within
these regions identify the correction targets. The resulting models,
candidate-identification metadata, decision criteria, and execution profiles
are made available to the runtime.

\parlabel{Constructing Correction Bindings.}
To associate each correction target with its feature-space context,
\ours{} organizes correction targets into \emph{Bindings}, which serve as the unit of specialized correction throughout the system. Each \emph{Binding} associates a \emph{Base} source class and correction target with a cluster in the VFM feature space, allowing the same source--target confusion to be handled differently across local feature contexts. \ours{} then prepares correction capabilities for these \emph{Binding}-specific contexts. We next describe the lightweight feature adaptation used as the first correction stage and the alternative visual correction available for additional correction.

\parlabel{Preparing Lightweight Adaptation.}
\ours{} first uses a lightweight correction stage to preserve most of the residual serving budget for samples that require additional processing. 
We implement this low-cost stage with feature adapters inspired by parameter-efficient adaptation~\cite{adapter, lada}: the VFM remains frozen, while a small learned module specializes its representation for a recurring correction target. Because the adapter operates on the representation produced by the VFM, \ours{} can reuse the feature already computed during \emph{Base} inference, avoiding an additional VFM feature-extraction pass for lightweight correction.
Each adapter applies a small residual transformation to this cached representation, which is then evaluated by the same frozen classifier to obtain a candidate correction. During offline preparation, \ours{} trains the adapters on labeled samples covering their correction targets and feature-space contexts, with only the adapter parameters updated. Reusing the representation keeps this first correction stage inexpensive, preserving serving resources for alternative visual correction when more processing is allocated to a sample.

\parlabel{Preparing Residual Decision Refinement.}
After lightweight adaptation, \ours{} can allocate additional processing to
samples selected for further examination~\cite{dvt}. For these samples, refinement
revisits the current prediction using additional image-derived features
together with the decision state accumulated so far. For each supported
\emph{Binding}, \ours{} prepares an alternative visual correction model that
evaluates the \emph{Binding}'s correction hypothesis in the context of the
\emph{Base} prediction and any prior correction decision. During offline
preparation, these models are trained on decision contexts representative of
those encountered during refinement, together with examples that encourage
preservation of correct decisions. This design allows \ours{} to invest
additional computation in selected samples while keeping lightweight
adaptation as the default first correction step.

\parlabel{Preparing Candidate-Identification State.}
A sample typically requires only a small subset of the correction capabilities prepared offline, so \ours{} also prepares the metadata needed to identify relevant \emph{Bindings} without executing every correction model. Using labeled offline data and the deployed VFM, \ours{} characterizes the \emph{Base} decisions and feature-space regions associated with each \emph{Binding} and stores compact descriptors of this information with the \emph{Bindings} catalog. During serving, the \emph{Base} prediction and cached VFM representation are compared against these descriptors to determine whether correction should be considered and, if so, which \emph{Bindings} become candidates for the sample. This candidate-identification step restricts subsequent correction to capabilities relevant to the observed \emph{Base} decision while avoiding exhaustive evaluation of the prepared \emph{Bindings}.

\parlabel{Preparing Correction-Decision Criteria.}
Executing a prepared correction model produces a candidate prediction or score, but this output does not directly determine whether the current prediction should be updated. \ours{} therefore prepares decision criteria that translate these model outputs into concrete correction actions at runtime.
Using labeled offline data, \ours{} determines the operating criteria under which a \emph{Binding} can introduce a correction, preserve an existing one, or leave the current prediction unchanged.
During serving, \ours{} evaluates each correction output against these pre-established criteria and updates the current prediction accordingly. 

\parlabel{Profiling Execution Cost.}
Whether a prepared correction capability can be used at runtime depends on whether its execution fits within the resource budget of the current inference window. \ours{} therefore profiles the execution time and GPU memory footprint of each correction capability before deployment~\cite{nestdnn}. The effective runtime cost can also depend on whether the required model state is already resident on the GPU, since a non-resident model incurs additional loading overhead~\cite{gemel}. During serving, \ours{} combines these profiles with the remaining inference time and available GPU memory to determine which correction work is predicted to fit within the current resource budget.

\subsection{Online Correction}
\label{sec:online-correction}

During serving, the optional correction work associated with each sample differs, while the residual resource budget available for that work varies across inference windows. \ours{} therefore treats online correction as resource-aware execution~\cite{nestdnn} over the capabilities prepared offline, adapting the amount of correction executed for each sample to the remaining inference time and GPU memory. Throughout this process, each sample maintains a valid current prediction, allowing \ours{} to stop additional correction when the current window cannot accommodate further work.

\parlabel{Base Inference and Correction Admission.}
\ours{} begins each inference window by executing the deployed VFM and shared classifier to obtain the \emph{Base} prediction and its softmax confidence per sample~\cite{uncertain, calibration}. To avoid spending correction resources on high-confidence decisions, \ours{} evaluates this confidence using the admission criterion prepared offline and filters out samples that do not enter the correction path. Samples that pass admission retain their VFM representations and proceed to candidate \emph{Bindings} identification.

\parlabel{Mapping Samples to Candidate Bindings.}
For each admitted sample, \ours{} uses its cached VFM representation to locate the sample within the feature-space clustering constructed offline. Each cluster is associated with \emph{Bindings} derived from the confusion patterns observed in that region, together with the correction capabilities prepared for those contexts. \ours{} therefore considers only the \emph{Bindings} associated with the sample's assigned cluster rather than all \emph{Bindings} compatible with its \emph{Base} class. This mapping directs each sample to correction capabilities prepared for a similar feature-space context while avoiding unnecessary execution of \emph{Bindings} specialized to other regions.

\parlabel{Lightweight Adapter Correction.}
For each candidate \emph{Binding} \(b\), \ours{} first evaluates its lightweight feature adapter using the VFM representation already cached during \emph{Base} inference. 
Let \(f\) denote this cached representation and \(A_b\) the feature adapter referenced by \emph{Binding} \(b\). The adapter produces a \emph{Binding}-specific representation \(f_b = f + A_b(f)\), where \(A_b(f)\) denotes the residual feature transformation learned for \emph{Binding} \(b\). 
This representation is passed through the same frozen classifier used for the \emph{Base} prediction, yielding a candidate prediction without an additional VFM forward pass. \ours{} then evaluates the resulting classifier output according to the \emph{Binding}'s correction criterion; a \emph{Binding} that satisfies this criterion produces a correction proposal for its target class, while otherwise that \emph{Binding} produces no proposal. Because multiple candidate \emph{Bindings} can produce proposals for the same sample, \ours{} composes these proposals before committing the lightweight correction result.

\parlabel{Refining Selected Decisions.}
For a sample selected for refinement, \ours{} invokes the alternative visual correction model associated with the relevant \emph{Binding}. Unlike the lightweight adapter, which reuses the cached VFM representation, the refinement model performs a new forward pass on the image and combines the resulting visual representation with the sample's current decision state. The refinement model is trained offline on decision contexts representative of those encountered during refinement, including cases that require a correction decision to change and examples that preserve an already correct decision. Its output is therefore a \emph{Binding}-specific score for updating the current decision rather than a standalone class prediction. \ours{} evaluates this score using the decision criterion prepared offline and updates or preserves the current prediction accordingly.

\parlabel{Scheduling Binding-Local Correction Pipelines.}
\ours{} organizes correction work around the \emph{Bindings} selected for admitted samples. 
For each \emph{Binding}, it maintains an ordered correction pipeline whose stages represent correction actions for the samples mapped to that \emph{Binding}, while the stage order preserves the per-sample dependencies between successive actions.
After a correction stage completes, its decision outcome determines whether the sample stops or becomes eligible for the next stage.
As these dependencies are satisfied, samples requiring the same \emph{Binding}-specific capability become ready independently and can be grouped for execution.
\ours{} schedules these ready groups by prioritizing earlier correction stages and using the \emph{Base} decision to rank competing work, while allowing other ready groups to proceed when higher-priority work is temporarily blocked by model availability.
This organization preserves the correction order required by each sample while exposing model-compatible work that can be executed together without imposing a global barrier across \emph{Bindings} or correction stages.
The execution order determines which optional work can fit within the inference window but does not change how correction outcomes are composed.

\parlabel{SLO-Aware Dispatch and Replanning.}
Scheduling determines which ready correction work is preferred, but priority alone does not determine whether that work can complete within the serving SLO~\cite{mark, adainf}. Before dispatching a correction action, \ours{} uses the remaining inference time and available GPU memory together with the execution profiles prepared offline to determine whether the action is predicted to fit within the current resource budget. As correction completes and model availability changes, \ours{} updates the remaining budget and reevaluates the remaining correction work. Actions that no longer fit are pruned together with their dependent work, and the affected samples retain their most recent valid predictions. In this way, \ours{} limits optional correction to work that is predicted to remain within the serving constraints of the current inference window.

\parlabel{Managing Model Residency and Prefetching.}
Keeping all \emph{Binding}-specific refinement models resident simultaneously can exceed the GPU memory available for optional correction. \ours{} therefore keeps the refinement model weights in pinned host memory and maintains a bounded cache of GPU-resident copies for the current correction workload~\cite{mocha}. Using the correction pipeline schedule to anticipate near-term demand, \ours{} asynchronously prefetches non-resident models from pinned host memory into the GPU cache, overlapping their transfer with currently executable correction work. Prefetching is bounded by the available GPU memory, and models associated with dispatched or near-term work are protected from eviction while other resident models can be replaced as the workload evolves. If a required model is still unavailable when its correction action becomes ready, independent ready work can continue until the 
transfer completes. This scheme bounds the memory consumed by 
correction while reducing model-loading stalls through overlap between transfer and useful computation.
\section{Evaluation}
\label{sec:evaluation}

\begin{figure}[t]
\centering
\includefigure[width=\linewidth]{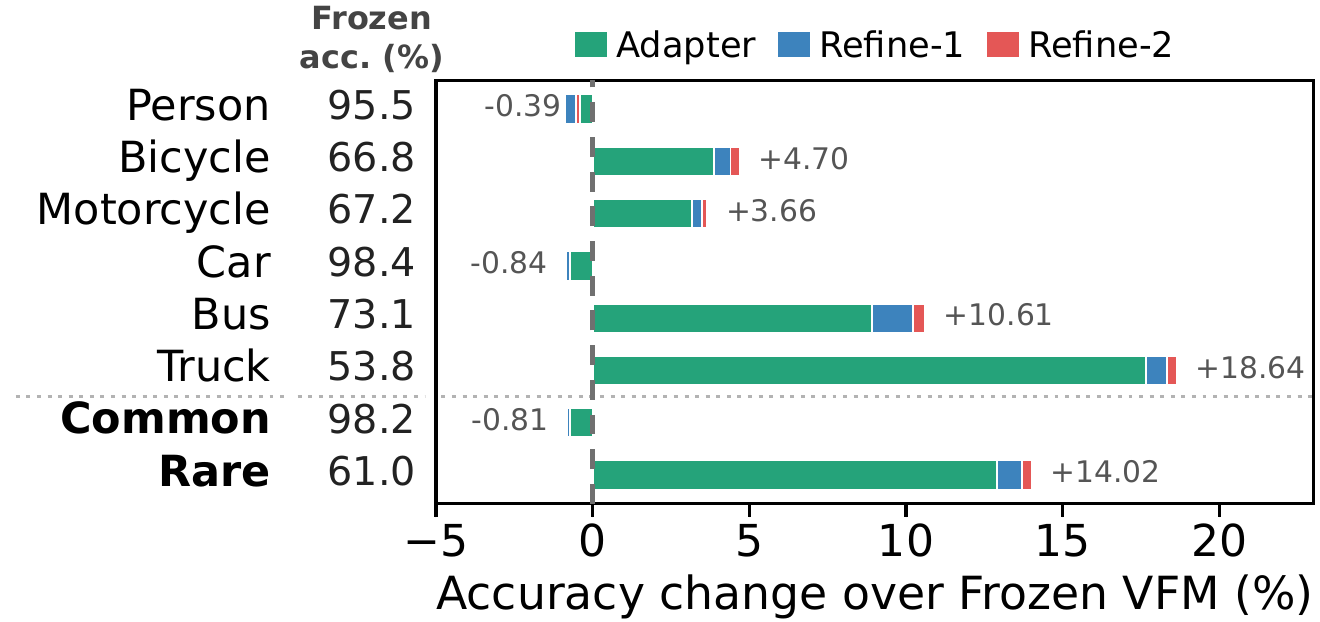}

\caption{
Per-class accuracy gains of \ours{} under single-stream serving,
aggregated over the four streams.
Left values show the accuracy of \textsf{Frozen}, and
the stacked bars decompose the
gains from the adapter and two refinement stages.
}
\label{fig:accuracy_change_decomposition}
\end{figure}

We evaluate \ours{} as a resource-aware correction layer on top of a frozen
VFM. Building on \autoref{sec:empirical-studies}, we ask whether \ours{} can
turn the GPU capacity left after VFM inference into higher rare-class
accuracy without disrupting the primary inference path. We measure the
accuracy gain and how much of it survives under stream concurrency
(\autoref{sec:eval-capability}) and under tighter compute and memory budgets
(\autoref{sec:eval-scaling}), then extend the same correction path to other
VFMs and devices (\autoref{sec:eval-generality}). Finally, we apply \ours{}
to an object-detection pipeline to test whether it works beyond
classification (\autoref{sec:eval-detection}).

\subsection{Methodology}
\label{sec:eval-methodology}

\parlabel{Workloads and Accuracy Metrics.} Unless otherwise noted, we use the
same platform, classification streams, and class definitions as
in~\autoref{sec:empirical-setup}. Serving proceeds in inference windows of 60
samples that arrive together, and each window must be processed within a 1~s
deadline before the next window arrives. We take the last prediction accepted
before each window's deadline and discard any correction that completes
afterwards. Rare-class accuracy is computed over all samples with rare ground-truth labels. To relate resource constraints to the amount of correction that
completes online, we also report the timely execution rate for the adapter
and the two refinement stages, defined as the fraction of issued requests
that complete before their window's deadline, regardless of whether the
correction changes the predicted label.

\parlabel{Scaling Interval.} The experiments that stress the compute and
memory budgets use a worst-case interval observed in our trace to have many
samples that need correction at once. We select the five-minute interval with
the highest rare-class confusion rate, defined as the fraction of samples
that \textsf{Frozen} misclassifies into a dominant confusion pair, using its
predictions and the ground-truth labels so that the choice does not depend on
any \ours{} result. All other experiments instead evaluate end-to-end
deployment behavior and therefore use the complete traces.

\subsection{Correction Effectiveness}
\label{sec:eval-capability}

\begin{figure}[t]
\centering
\includefigure[width=\linewidth]{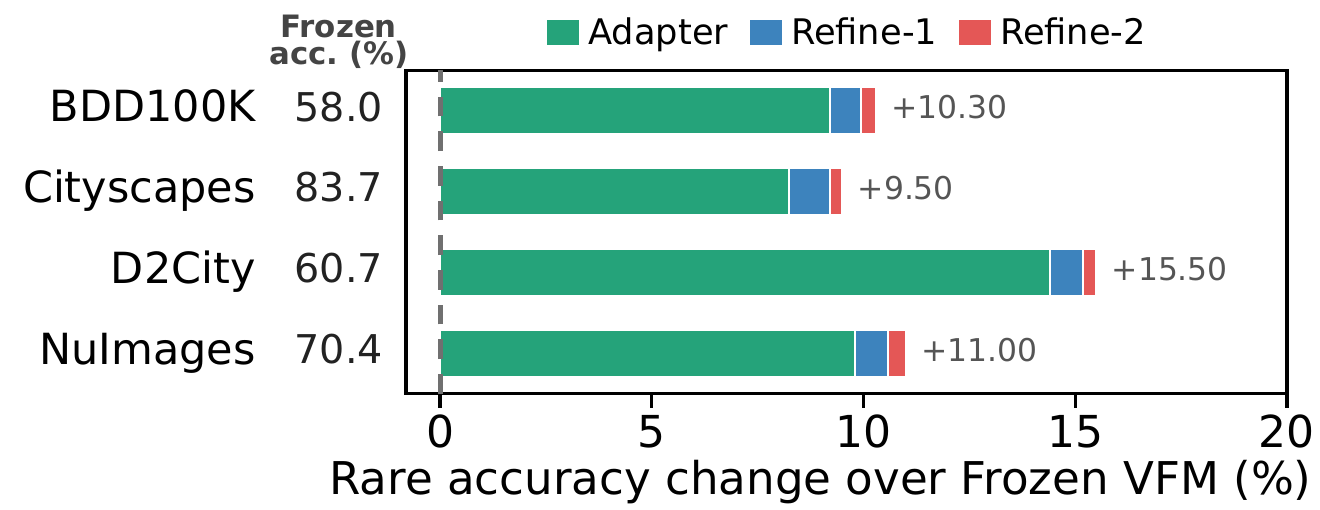}

\caption{
Rare-class accuracy gains of \ours{} on each of the four deployment
streams under single-stream serving.
}
\label{fig:dataset_accuracy_change_decomposition}
\end{figure}

We first evaluate how much \ours{} improves rare-class accuracy when compute
and memory suffice to execute every eligible correction stage.
\autoref{fig:accuracy_change_decomposition} reports the accuracy gain over
\textsf{Frozen}, aggregated over the four streams, as the adapter and the two
refinement stages are progressively enabled. The adapter alone raises
\textsf{Truck} accuracy from 53.8\% to 71.5\%, and the
two refinement stages add 0.7\% and 0.3\%. Meanwhile,
\textsf{Car} and \textsf{Person} lose 0.8\% and 0.4\%, respectively, so the
rare-class gains come with little degradation on the common classes.
\autoref{fig:dataset_accuracy_change_decomposition} breaks the same
measurement down by stream. Although the rare-class accuracy of
\textsf{Frozen} differs from stream to stream, the average rare-class
accuracy improves by 9.5\% to 15.5\% across the four
streams, with the adapter again providing most of the gain.

The gain therefore diminishes with correction depth: the cheapest stage
recovers most of the accuracy, and each deeper stage adds less while covering
harder samples that earlier stages could not fix. This is the property that
\autoref{sec:opportunities} identifies as necessary for partial correction to
be useful, and \autoref{sec:eval-scaling} shows that \ours{} can skip the
deeper stages under a tight budget and still retain most of the accuracy
gain.

\begin{figure}[t]
\centering
\includefigure[width=\linewidth]{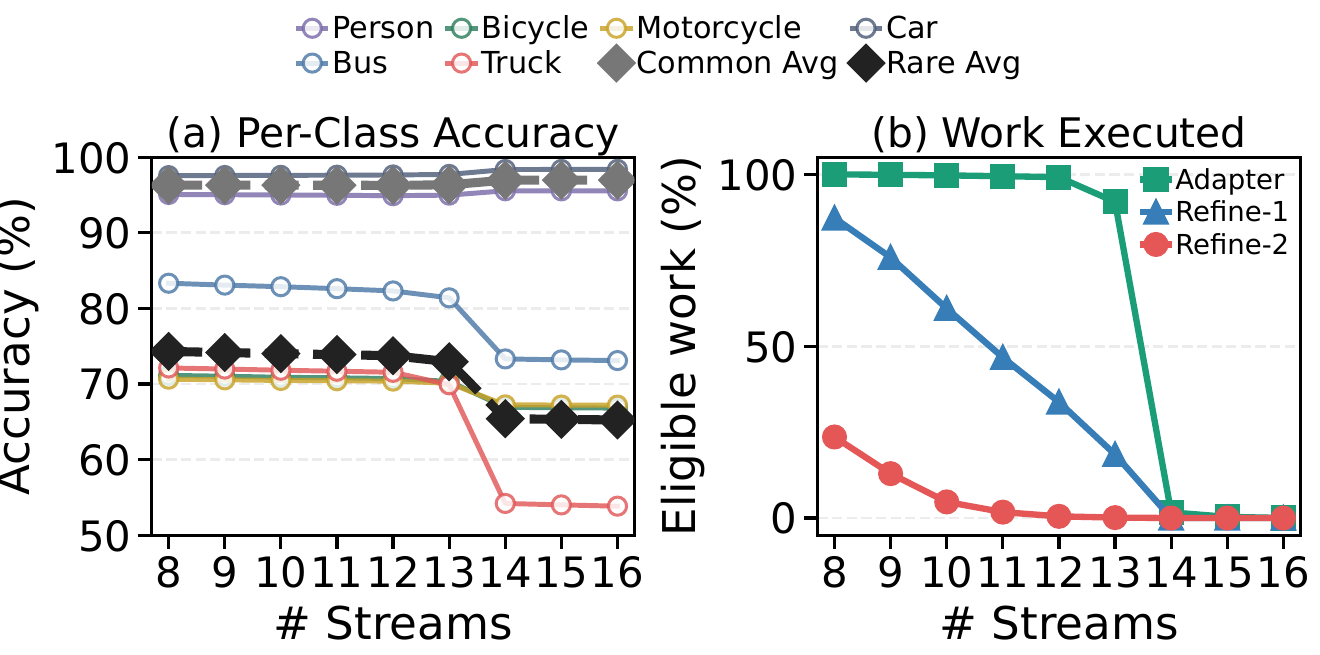}

\caption{
Scaling of \ours{} with increasing stream concurrency. (a) Per-class accuracy,
including the common-class and rare-class averages. (b) Timely execution rate
at each correction stage.
}
\label{fig:stream_scaling_per_class_accuracy}
\end{figure}

\parlabel{Stream Scaling.} We next co-locate more streams on a fixed server
capacity, which is how contention arises in deployment.
\autoref{fig:stream_scaling_per_class_accuracy}(a) shows that the rare-class
gain barely changes as concurrency increases up to thirteen streams, so
\ours{} retains most of its benefit even as the headroom per stream initially
shrinks. The timely execution rates in
\autoref{fig:stream_scaling_per_class_accuracy}(b) explain why: the adapter
continues to execute almost all of its eligible work, while the deeper
refinement stages are the first to be dropped. At fourteen streams, the
remaining headroom can no longer sustain even the adapter, and the gain falls
close to zero.

The window miss rate, the fraction of inference windows whose final
prediction is produced after the 1\,s deadline, stays at 1.0\% or below
through eight streams, yet rises to 10.7\% at twelve
(\autoref{tab:eval-slo}). These misses come from dynamic contention among
co-located streams, since MPS~\cite{nvidia_mps} does not isolate GPU resources between them. 
\ours{} drops optional correction work before the deadline, but cannot
shorten the VFM inference that every stream must run: at fourteen streams the
serving path alone can no longer keep pace with incoming work. We show below
that a stronger guarantee of resource share substantially reduces the window
miss rate.

\begin{table}[t]
\centering
\captionsetup{width=\columnwidth,skip=3pt}

\caption{
SLO behavior under pressure. A miss occurs when the prediction exceeds the 1\,s deadline.
}
\label{tab:eval-slo}

\renewcommand{\arraystretch}{1.02}

\begin{tabular*}{\columnwidth}{@{\extracolsep{\fill}}lrrrrr@{}}
\toprule

\textbf{Streams (N)}
& 8 & 10 & 12 & 14 & 16 \\
\textit{Miss (\%)}
& 1.00 & 5.63 & 10.67 & 42.88 & 97.88 \\

\midrule

\textbf{Compute (GPU)}
& 100\% & 50\% & 25\% & 12\% & 6\% \\
\textit{Miss (\%)}
& 0 & 0 & 0 & 0.67 & 2.00 \\

\midrule

\textbf{Memory (GB)}
& 12 & 6 & 3 & 1.5 & -- \\
\textit{Miss (\%)}
& 0 & 0 & 0 & 0 & -- \\

\bottomrule
\end{tabular*}

\end{table}

\begin{figure}[t]
\centering
\includefigure[width=\linewidth]{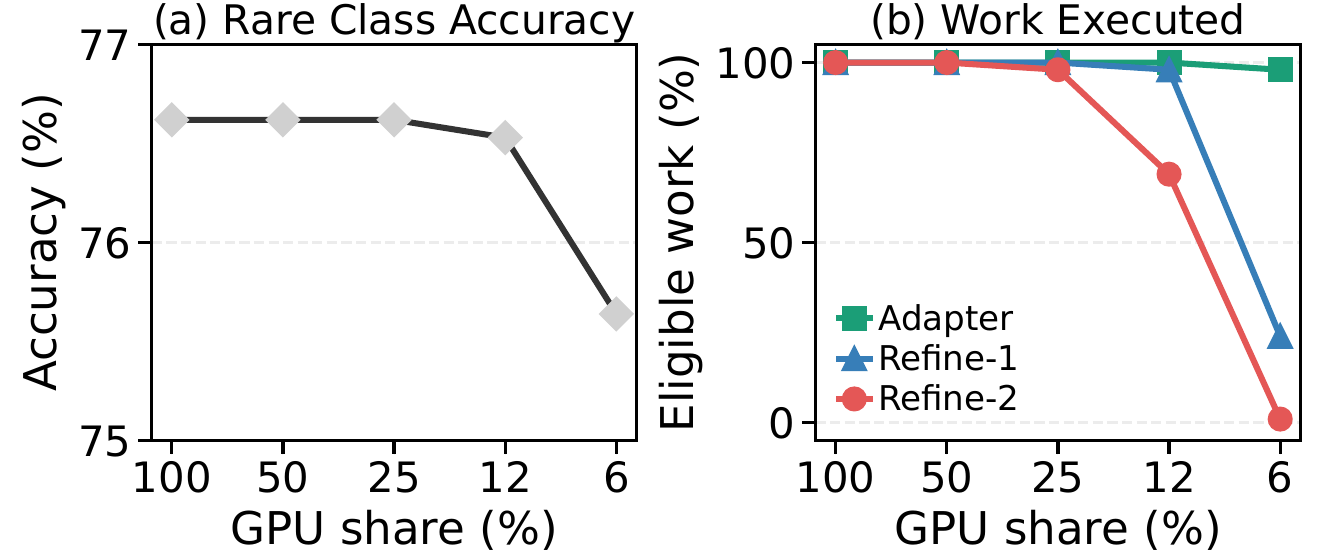}

\caption{
Compute scaling of \ours{} under decreasing GPU allocation. (a) Rare-class
accuracy and (b) timely execution rate at each correction stage.
}
\label{fig:compute_resource_scaling}
\end{figure}

\subsection{Resource Sensitivity}
\label{sec:eval-scaling}

Stream concurrency tightens the compute and memory budgets together, so we
next vary each budget alone on the scaling interval to identify where it
starts to limit correction.

\parlabel{Compute.} \ours{} preserves nearly all of its accuracy gain with
only a fraction of the fully provisioned compute.
\autoref{fig:compute_resource_scaling} shows that rare-class accuracy stays near
76.6\% as the per-stream MPS allocation is reduced from 100\% to 12\%. The
timely execution rates show the same pattern as before: the deepest
refinement stage is dropped first, while the adapter keeps executing. At the
most constrained 6\% allocation, rare-class accuracy falls slightly to
roughly 75.6\%, as execution of the refinement stages collapses. The adapter
still executes most of its eligible work, so even this setting retains much
of the accuracy gain.

With a dedicated allocation, \ours{} keeps window misses low across a wide
range of GPU shares. As shown in \autoref{tab:eval-slo}, no misses occur down
to a 25\% allocation, while the miss rate is only 0.7\% at 12\% and 2.0\% at
6\%. This is a stark contrast with the MPS-based stream scaling results,
where the miss rate already reaches 10.7\% at twelve streams. Stricter
isolation, such as MIG~\cite{nvidia_mig}, would therefore help \ours{} sustain
correction at higher concurrency.

\begin{figure}[t]
\centering
\includefigure[width=\linewidth]{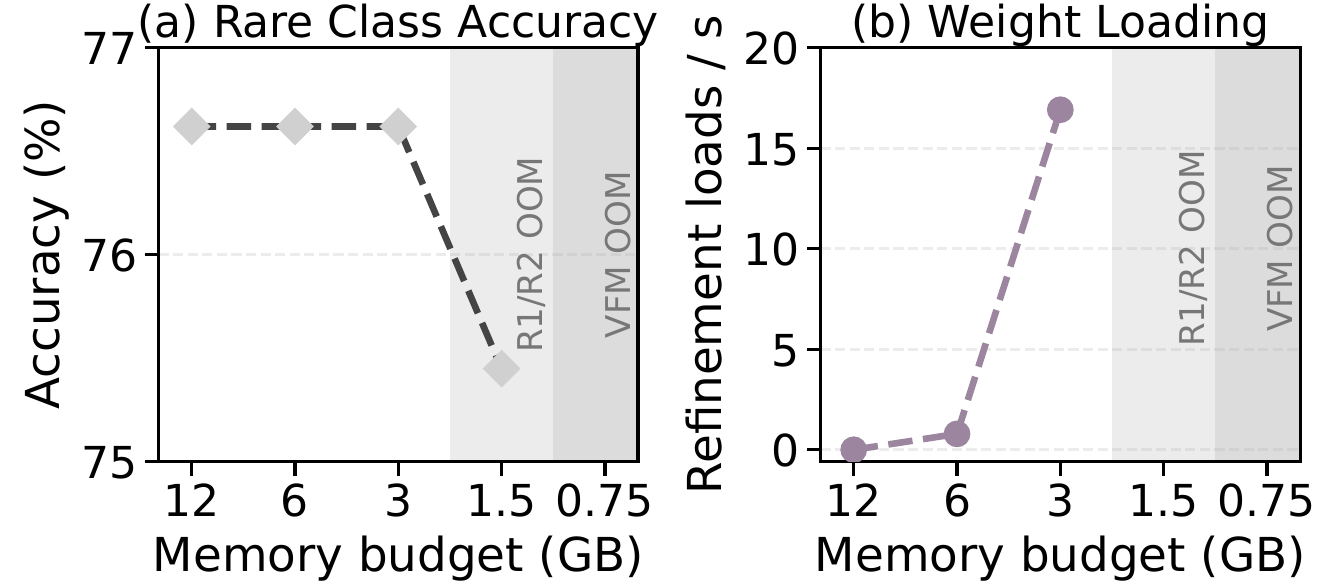}

\caption{
Memory scaling of \ours{} under decreasing GPU memory budgets.
(a) Rare-class accuracy and (b) the rate of refinement-model loads from host memory.
The shaded regions mark budgets where the refinement models (R1/R2) or
the VFM itself no longer fit.
}
\label{fig:eval-memory-scaling}
\vspace{-1pt}
\end{figure}

\parlabel{Memory.} \ours{} does not require the complete set of correction
models to stay resident on the GPU. \autoref{fig:eval-memory-scaling}(a)
shows that rare-class accuracy stays near 76.6\% as the per-stream memory
budget falls from 12~GB to 3~GB. This holds even though refinement-model
loads from host memory rise sharply at 3~GB
(\autoref{fig:eval-memory-scaling}(b)), since prefetching overlaps these
loads with ongoing correction work. GPU residency can therefore be
provisioned well below the footprint of the complete correction set.
Prefetching, however, cannot help once the budget is too small to hold the
refinement models at all. At 1.5~GB, only the adapter can run, yet rare-class
accuracy drops only slightly to roughly 75.5\%, since the adapter carries
most of the gain. At 0.75~GB, even the VFM inference stack no longer fits in GPU memory, causing OOM failure.
Window misses stay at 0\% from 12~GB down to 1.5~GB
(\autoref{tab:eval-slo}), so the loads that prefetching absorbs add little 
deadline pressure.

\begin{figure}[t]
\centering
\includefigure[width=\linewidth]{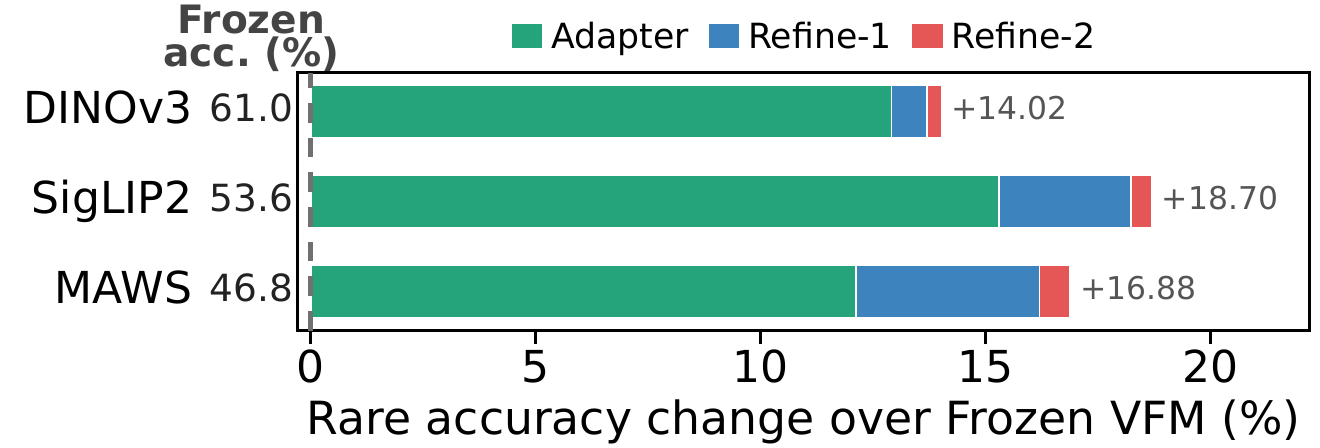}
\caption{
Rare-class accuracy gains of \ours{} across three VFM backbones.
}
\label{fig:rare_class_accuracy_across_three_VFMs}
\end{figure}

\subsection{Generality and Alternative Designs}
\label{sec:eval-generality}

\parlabel{Different VFMs.} \ours{} improves rare-class accuracy with other
VFMs as well.
\autoref{fig:rare_class_accuracy_across_three_VFMs} shows that rare-class
accuracy rises by 14.0\% with DINOv3-S, by 18.7\% with SigLIP2-B~\cite{siglip2}, and by 16.9\% with MAWS-B~\cite{MAWS}, from 61.0\%, 53.6\%, and 46.8\%, respectively. Each gain is measured against each VFM's own uncorrected configuration, with correction models prepared for its own feature space. As in \autoref{sec:eval-capability}, the adapter provides most of each gain, while the refinement stages add smaller increments whose size varies across the three VFMs.

\begin{table}[b]
\centering
\captionsetup{width=\columnwidth,skip=3pt}

\caption{
Comparison with alternative correction designs.
}
\label{tab:alternative-correction}

\setlength{\tabcolsep}{3pt}

\begin{tabular*}{\columnwidth}{@{\extracolsep{\fill}}lcccc@{}}
\toprule
\textbf{Correction Design}
& \textbf{Prototype}
& \textbf{kNN}
& \textbf{MLP}
& \textbf{\ours{}} \\
\midrule
\textbf{Rare-class gain (\%)}
& +3.4
& +4.5
& +6.0
& +14.0 \\
\bottomrule
\end{tabular*}
\end{table}

\parlabel{Edge Deployment.} We next move the same correction path to Jetson Orin NX~\cite{jetson_orin_nx}, whose compute envelope is far tighter than that of the RTX 3090 GPU used so far. 

\begin{figure}[t]
\centering
\includefigure[width=\linewidth]{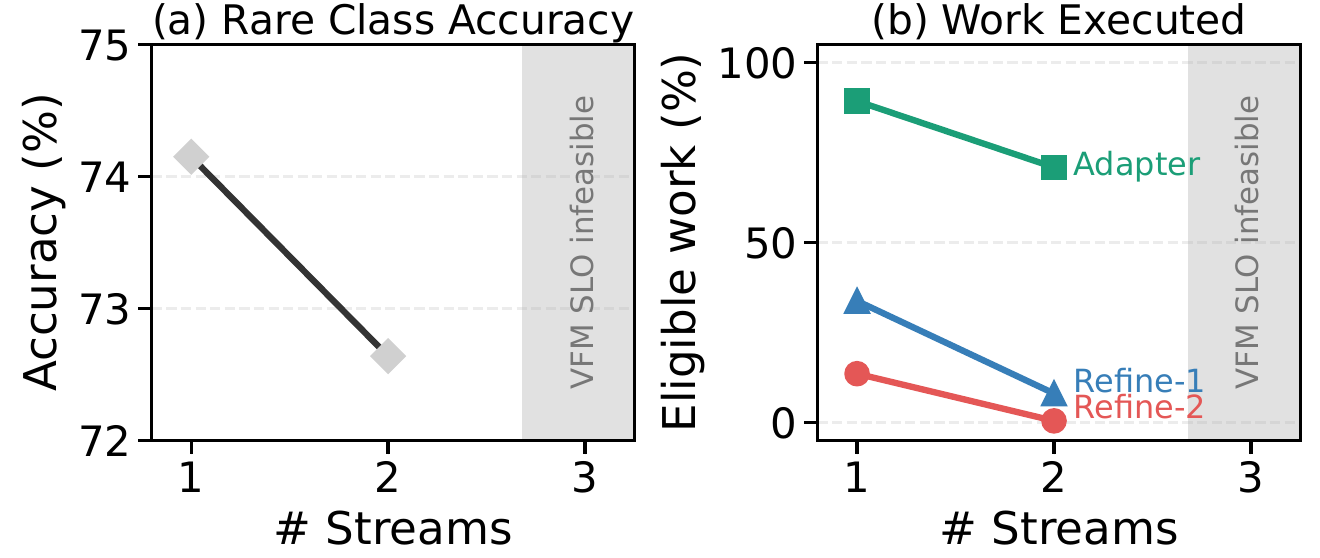}
\caption{
Scaling of \ours{} on Jetson Orin NX.
(a) Rare-class accuracy and (b) timely execution rate at each correction stage.
The shaded region marks three streams, where VFM inference alone exceeds
the deadline.
}
\label{fig:jetson_orin_nx}
\vspace{-5pt}
\end{figure}

To match that envelope, we replay each stream at 30 samples/s and increase concurrency from one to three streams. As \autoref{fig:jetson_orin_nx} shows, rare-class accuracy falls only from 74.2\% to 72.6\%, as \ours{} again skips the deeper stages first. The adapter's timely execution rate decreases from 89.3\% to 70.9\%, whereas the two refinement stages fall from 33.8\% and 13.6\% to 8.1\% and 0.5\%. The Jetson thus follows the same pattern observed on RTX 3090, despite its much smaller capacity. 
What limits this device is VFM inference itself rather than correction. At three concurrent streams, VFM inference alone misses the 1\,s deadline for 76.4\% of inference windows, even with all correction disabled. We therefore treat two streams as the maximum feasible operating point on this device.

\parlabel{Alternative Correction Designs.} We next compare \ours{} with
simpler correction designs to determine whether its rare-class gains can be
recovered without \emph{Binding}-specific staged correction. We evaluate a prototype
classifier~\cite{prototype_classifier}, feature-space kNN~\cite{feature_space_knn, dknn}, and a shared MLP~\cite{shared_mlp} under the
same common-class preservation constraint as \ours{}. As shown in
\autoref{tab:alternative-correction}, the three alternatives improve
rare-class accuracy by 3.4\%, 4.5\%, and 6.0\%, respectively, whereas \ours{}
achieves 14.0\%. While empirical, this comparison supports the design choice
we make in this paper.

\subsection{Beyond Classification: Object Detection}
\label{sec:eval-detection}

The experiments so far evaluate \ours{} on pre-cropped objects, but deployed pipelines typically run detection. Here, the same rare-class confusions appear: a truck that is localized correctly but recognized as a car is still a rare-class error. Detection also changes the correction workload, since each frame yields multiple objects whose recognition may need correction within the same deadline.

\parlabel{Setup.} We use a frozen DETR-based detector~\cite{DINO-DETR} that
predicts boxes and classes directly from the full frame. \ours{} runs after
the detector has completed its normal selection pipeline, so it changes
neither which objects are detected nor where their boxes are. It revises only
the predicted class of each detection, through the same admission criterion,
adapter, and refinement stages as in classification. We evaluate on the
original full frames of the online partitions and report AP$_{50}$, which
measures detection accuracy for each class, and its mean over the four rare
classes (rare-class AP).

\begin{figure}[t]
\centering
\includefigure[width=\linewidth]{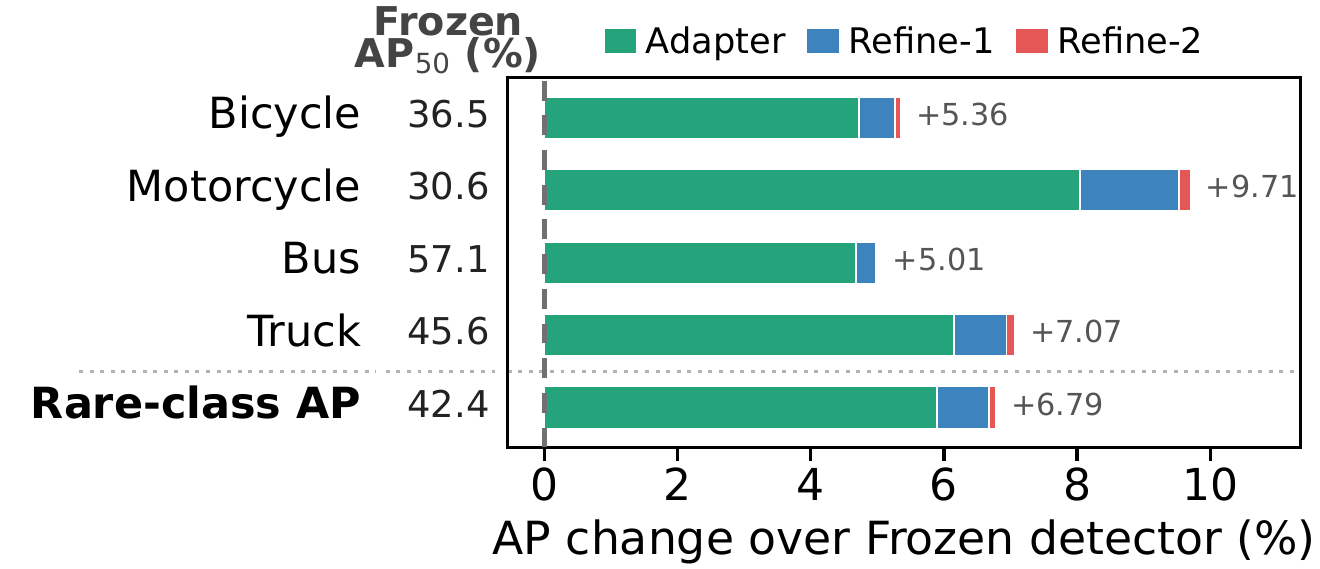}
\caption{
Rare-class AP gain of \ours{} on object detection.
}
\label{fig:eval-detection}
\end{figure}

\parlabel{Results.} \autoref{fig:eval-detection} shows that \ours{} raises
rare-class AP from 42.4\% to 49.2\%, with per-class gains of 5.4\% for
\textsf{Bicycle}, 9.7\% for \textsf{Motorcycle}, 5.0\% for \textsf{Bus}, and
7.1\% for \textsf{Truck}. As in classification, the adapter provides most of each
gain and the refinement stages add smaller increments. 
Because localization is left untouched, these gains come entirely from correcting
recognition errors.

\section{Discussions and Limitations}
\label{sec:limitations}

\parlabel{Multi-GPU Deployments.} Our evaluation places all streams on a
single GPU, where correction competes with inference for the same device.
\autoref{sec:eval-scaling} shows that \ours{} sustains correction far better under a dedicated allocation than under MPS sharing, which suggests two directions for \ours{} on a multi-GPU server: streams can be partitioned across GPUs to mitigate per-GPU contention, or correction models can be placed on a separate device so that one stream's correction no longer contends with another stream's inference. Both require modifications to the budget-aware scheduler, but neither changes the correction pipeline itself.

\parlabel{Sharing Correction Models across Streams.} \ours{} prepares
correction models offline from public data, so once constructed, the adapter and refinement models could be applied to every stream that shares the VFM and a class set. Our multi-stream experiments already reuse a single correction pipeline across all co-located streams. A server can therefore add a stream without going through offline preparation, although sharing one copy of the models across MPS instances would require additional system support.

\parlabel{Limits of Offline Preparation.} \ours{} targets dominant confusion pairs that can be identified before deployment, so it is less effective when mispredictions spread evenly across many class pairs, when previously unseen classes appear, or when the deployment distribution drifts far enough to invalidate the current correction pipeline. These cases are where continuous learning still has a role. Combining periodic model updates with the selective correction of \ours{} is a complementary direction that we leave for future work.

\section{Related Work}

\ours{} builds on the empirical comparison in~\autoref{sec:empirical-studies} against retraining methods that allocate resources and schedule retraining for high average accuracy~\cite{ekya, recl, eomu}, reuse models for scalability~\cite{recl, mocha}, or apply incremental updates to harvest spare time~\cite{adainf}, and we do not repeat it here.

\parlabel{Edge Video Analytics Systems.} A number of systems optimize video analytics pipelines under resource constraints without addressing data drift. Chameleon~\cite{chameleon} adapts configurations such as resolution and frame rate in real time to balance accuracy and compute cost. Reducto~\cite{reducto} performs on-camera filtering to eliminate redundant frames before they reach the edge server. DDS~\cite{dds} uses server-side DNN feedback to guide camera encoding and streaming decisions, optimizing bandwidth without sacrificing inference accuracy. Gemel~\cite{gemel} reduces GPU memory pressure by merging architecturally similar layers across co-located models. These systems all assume a stationary data distribution and do not account for accuracy degradation due to data drift, which is a first-order concern in long-running deployments.

\parlabel{Continual Learning on the Edge.} Continual learning trains models
on sequential tasks, where each task introduces new classes or distribution
shifts, while seeking to retain previously learned
knowledge~\cite{gepperthH16}. For edge deployments, Kwon et al. empirically
analyzed rehearsal and regularization methods to characterize their
cost-accuracy trade-offs across the system stack~\cite{kwon2021exploring}.
Miro~\cite{miro} reduces energy consumption by dynamically tuning training
configurations for each new task. SparCL~\cite{sparcl} enables sparse
continual learning for CNNs on mobile devices, and REP~\cite{rep} reduces the
computation cost of ViT-based continual learning on edge devices. While VFMs
are widely adopted for various vision tasks~\cite{sam},
their implications for edge video analytics remain largely unexplored.

\parlabel{Other Adaptation Techniques.} Parameter-efficient methods adapt a frozen backbone with a few trainable weights, such as adapters~\cite{adapter}, LoRA~\cite{lora}, and visual prompt tuning~\cite{vpt}. These methods learn a single representation shared by all classes, whereas \ours{} prepares a  correction model for each confusion pair and feature-space group, which lets it apply correction only where rare-to-common misclassifications concentrate. Test-time adaptation updates the model while it serves the stream~\cite{tent, cotta}, which places gradient computation and a label-free loss on the serving path. \ours{} reuses the frozen feature and does no online update. 

\section{Conclusion}
\label{sec:conclusion}
Across four driving datasets, a frozen compact VFM achieves higher accuracy than retraining-based CL systems while using 6--47\% of the GPU time, but this average masks large rare-class accuracy gaps. \ours{} uses the GPU capacity left after VFM inference for staged correction. A lightweight adapter operates on cached VFM features, and costlier refinement stages run only when the latency and memory budgets allow. \ours{} improves rare-class accuracy by 14.0\% with less than 1\% common-class degradation, and remains effective across three VFM backbones, on a Jetson Orin NX, and for object detection.

\footnotesize
\bibliographystyle{plain}
\bibliography{References}

\end{document}